\documentclass[10pt,final,journal]{IEEEtran}

\usepackage{cite}
\usepackage{amsmath,amssymb,amsfonts}
\usepackage{algorithmic}
\usepackage{graphicx}
\usepackage{textcomp}
\usepackage{xcolor}
\usepackage{array,multirow}
\def\BibTeX{{\rm B\kern-.05em{\sc i\kern-.025em b}\kern-.08em
    T\kern-.1667em\lower.7ex\hbox{E}\kern-.125emX}}

\begin{document}
%
\title{First Demonstration of the Korean eLoran Accuracy in a Narrow Waterway Using Improved ASF Maps}
%
%
%

        
\author{
Woohyun~Kim, Pyo-Woong~Son, Sul~Gee~Park, Sang~Hyun~Park, and~Jiwon~Seo,~\IEEEmembership{Member,~IEEE}
\thanks{Manuscript received June 00, 2021; }
\thanks{This research was conducted as a part of the projects titled ``Development of enhanced Loran (eLoran) system [PMS4020]'' and ``Development of integrated R-Mode navigation system [PMS4440]'' funded by the Ministry of Oceans and Fisheries, Republic of Korea. This research was also supported in part by the Institute of Information and Communications Technology Planning and Evaluation (IITP) grant funded by the Korea government (KNPA) (2019-0-01291).
}
\thanks{Authors' addresses: W.~Kim and J.~Seo are with the School of Integrated Technology, Yonsei University, Incheon 21983, Republic of Korea, E-mail: (crimy00@yonsei.ac.kr; jiwon.seo@yonsei.ac.kr); P.-W.~Son, S.~G.~Park, and S.~H.~Park are with the Korea Research Institute of Ships and Ocean Engineering, Daejeon 34103, Republic of Korea, E-mail: (pwson@kriso.re.kr; sgpark@kriso.re.kr; shpark@kriso.re.kr). \it{(Corresponding author: Jiwon Seo.)}}
}

%
%

%
\markboth{IEEE Transactions on Aerospace and Electronic Systems,~Vol.~00, No.~0, June~2021}%
{Kim \MakeLowercase{\textit{et al.}}: First Demonstration of the Korean eLoran Accuracy in a Narrow Waterway Using Improved ASF Maps}

\maketitle

\begin{abstract}
The vulnerabilities of global navigation satellite systems (GNSSs) to radio frequency jamming and spoofing have attracted significant research attention. In particular, the large-scale jamming incidents that occurred in South Korea substantiate the practical importance of implementing a complementary navigation system. This letter briefly summarizes the efforts of South Korea to deploy an enhanced long-range navigation (eLoran) system, which is a terrestrial low-frequency radio navigation system that can complement GNSSs. After four years of research and development, the Korean eLoran testbed system has been recently deployed and is operational since June 1, 2021. Although its initial performance at sea is satisfactory, navigation through a narrow waterway is still challenging because a complete survey of the additional secondary factor (ASF), which is the largest source of error for eLoran, is practically difficult in a narrow waterway. This letter proposes an alternative way to survey the ASF in a narrow waterway and improve the ASF map generation methods. Moreover, the performance of the proposed approach was validated experimentally. 
\end{abstract}


%
\IEEEpeerreviewmaketitle

\section{Introduction}
\label{sec:Intro}
%
%
%
%
\IEEEPARstart{T}{he} vulnerabilities of global navigation satellite systems (GNSSs), such as GPS of the U.S. and Galileo of Europe, to radio frequency jamming and spoofing are well known. Therefore, potential complementary navigation systems have been actively studied. Among these systems, the enhanced long-range navigation (eLoran) \cite{Pelgrum06, Williams13}, which is a terrestrial low-frequency radio navigation system, has demonstrated potential as an effective complementary system because eLoran has very different failure modes than the GNSSs.

After experiencing a series of high-power GPS jamming incidents, South Korea decided to deploy an eLoran testbed system as part of the first phase of its eLoran program. This testbed covers the northwestern part of the country. The performance of the eLoran testbed, which has been operational since June 1, 2021, is currently being evaluated for its feasibility for a nationwide eLoran system, which is the second phase of the program.

The demonstrated accuracy of the eLoran testbed during the initial tests on the sea fulfilled the 20-m (95\%) accuracy requirement in the request for proposal (RFP) of the project. However, providing a high accuracy in a narrow waterway is challenging. 
An additional secondary factor (ASF), which is the signal propagation delay due to the land path \cite{Lo09}, is the largest source of error for eLoran. To mitigate the ASF errors, it is necessary to survey ASF values in the service area and generate ASF maps for users before the provision of eLoran service. 
ASF surveys in narrow waterways are practically difficult because only a small ship with high angular motions can be used for the survey; thus, ASF measurements are noisier than in the case of a large and stable ship. Furthermore, the ASF variation at the land-sea boundary is significantly large and causes non-negligible cross-track ASF variations. 

In this letter, we propose a simple cross-track ASF survey in addition to a sparse along-track survey in a narrow waterway to improve the quality of ASF maps while minimizing the time and effort required for ASF surveys. In addition, novel ASF map generation methods that effectively incorporate the trend of cross-track ASF variations are proposed. The proposed survey and ASF map generation methods were validated using field tests in the Ara Waterway, Incheon, Korea.

\section{Development of Korean eLoran Testbed System}
The eLoran system attracted significant attention in South Korea because of the exhibited vulnerability of GPS during repeated high-power GPS jamming from North Korea since 2010. However, since the initial announcement of the South Korean eLoran program in 2013 \cite{Seo13}, the procurement process of the necessary eLoran equipment through international competitive bidding was not successful. (Detailed milestones of the Korean eLoran program are given in Table \ref{table:eLoran_program}.) Consequently, the Ministry of Oceans and Fisheries (MOF) of South Korea lost the momentum to continue the plan to purchase and deploy an entire eLoran system from the market. Thus, the MOF decided to transform the eLoran program into a research and development (R\&D) project, which includes the development of an eLoran transmitter and operational software \cite{Son20}. 

\begin{table}[!t]
\renewcommand{\arraystretch}{1.3}
\caption{Milestones of the Korean eLoran Program}
\label{table:eLoran_program}
\centering
\begin{tabular}{|p{2.1cm}||p{5.5cm}|}
\hline
Aug. 23--26, 2010 & First North Korean jamming incident occurred.\\
\hline
Mar. 4--14, 2011 & Second North Korean jamming incident occurred.\\
\hline
Oct. 19, 2011 & Korean eLoran program was initiated.\\
\hline
Apr. 28--May 13, 2012 & Third North Korean jamming incident occurred.\\
\hline
Apr. 23, 2013 & Korean eLoran program was internationally announced based on the initial design of a nationwide system.\\
\hline
Nov. 2013--Jan. 2014 & Three rounds of bidding failed due to a lack of bidders or unacceptable proposals.\\
\hline
Apr. 14, 2014 & Revised plan with a two-phase approach was internationally announced.\\
\hline
Oct. 2014--Jun 2015 & The contract was awarded, but the awarded contract was breached due to unpublicized issues of the consortium.\\
\hline
Jan. 15, 2016 & The RFP for research and development of Korean eLoran testbed system was released.\\
\hline
Mar. 31--Apr. 5, 2016 & Fourth North Korean jamming incident occurred.\\
\hline
May 27, 2016 & The KRISO consortium was awarded the R\&D project.\\
\hline
Jun. 1, 2021 & The Korean eLoran testbed system was announced to be operational.\\  
\hline
\end{tabular}
\end{table}

This R\&D project was awarded to the consortium led by the Korea Research Institute of Ships and Ocean Engineering (KRISO) in 2016.
As a separate project, the MOF synchronized the timing of the existing Loran-C \cite{Pu21, Yuan20, Qiu10, Son18} transmitters in Pohang and Gwangju to the coordinated universal time (UTC) because at least three UTC-synchronized transmitters are required to calculate the two-dimensional (2-D) position of the user. 
The eLoran testbed system developed by the KRISO consortium covering the northwestern region of South Korea has been operational since June 1, 2021, which took more than ten years from the first North Korean GPS jamming in 2010. On achieving a satisfactory performance, South Korea plans to deploy two more eLoran transmitters and become the first country with a nationwide eLoran system.

\section{Proposed ASF Survey and ASF Map Generation Methods for a Narrow Waterway}
\label{sec:Methods}

\begin{figure}
  \centering
  \includegraphics[width=0.7\linewidth]{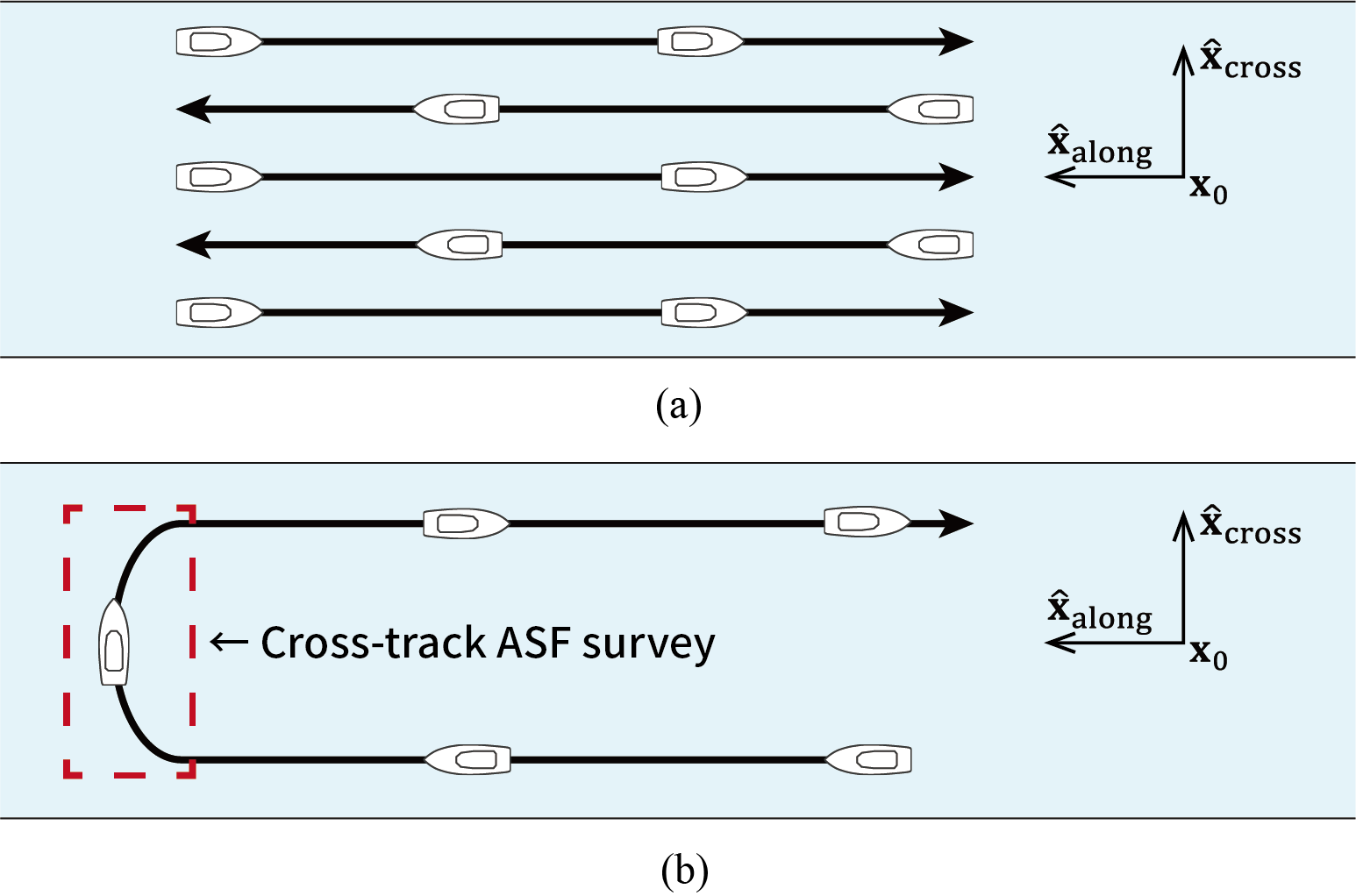}
  \caption{Comparison of (a) a conventional ASF survey method and (b) the proposed survey method for a narrow waterway. The $\mathbf{x}_0$ represents the origin of the local coordinate system located at the center of the waterway, and $\hat{\mathbf{x}}_\mathrm{along}$ and $\hat{\mathbf{x}}_\mathrm{cross}$ represent unit vectors in the along-track and cross-track directions, respectively.}
  \label{fig:SurveyComp}
\end{figure}

The operational performance of the eLoran testbed  is currently under evaluation. One of the challenges is to improve the positioning accuracy in a narrow waterway. As explained in Section \ref{sec:Intro}, it is difficult to survey the ASF in a narrow waterway. Considering the cross-track ASF variations in a waterway, a ship needs to survey the same along-track location multiple times with different cross-track points, as illustrated in Fig. \ref{fig:SurveyComp}(a), which requires significant time and effort. Instead, we propose a cross-track survey by turning the survey vessel in addition to a sparse along-track survey, as illustrated in Fig. \ref{fig:SurveyComp}(b). 

The idea here is to extract a cross-track ASF trend through a simple cross-track survey and apply the obtained trend to the region where only a sparse along-track survey was performed. 
To implement this idea, two questions should be answered: 1) How do you extract the cross-track ASF trend from noisy cross-track ASF survey data? 2) How do you utilize the cross-track trend to generate improved ASF maps for a narrow waterway?
These questions are addressed in the following subsections.

\subsection{Outlier Removal and Smoothing Spline Fit to Obtain Cross-Track ASF Trend}
\label{sec:ObtainAsfTrend} 

\begin{figure}
  \centering
  \includegraphics[width=0.7\linewidth]{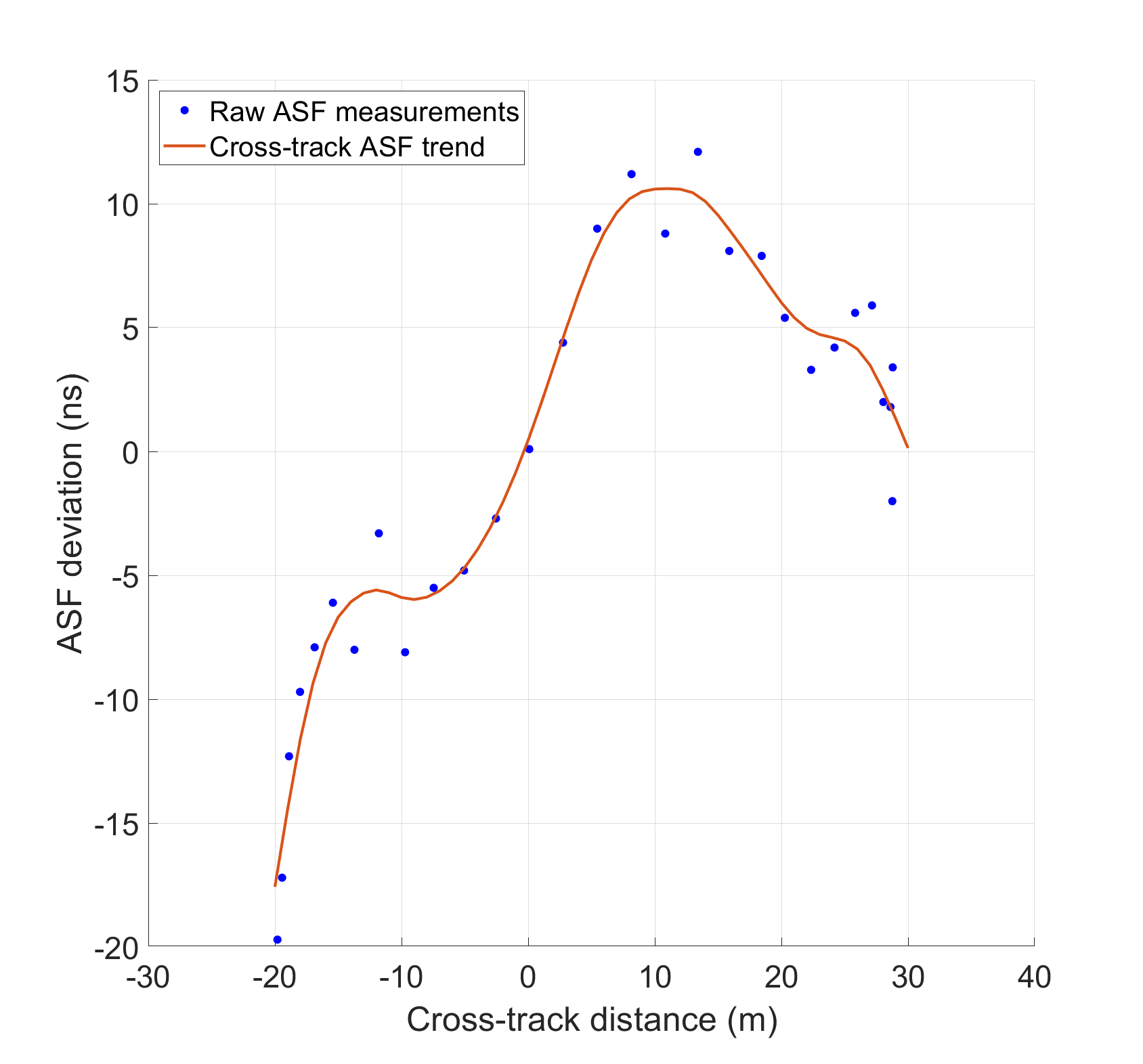}
  \caption{Cross-track ASF measurement deviations with respect to the ASF at the center of the waterway. Cross-track ASF trend was obtained as a cubic smoothing spline fit to the outlier-removed data. These ASF data were obtained based on the signals from the newly deployed eLoran transmitter (9930V) in Incheon, Korea.}
  \label{fig:ASFdeviations}
\end{figure}

In a narrow waterway, only a small vessel can be used for ASF survey, but the ASF measurements in this case are significantly noisier than the case of a large vessel because of its greater angular motions and hull vibrations. The actual ASF measurements during our cross-track survey in the red box of Fig. \ref{fig:SurveyComp}(b) have noticeable outliers that do not follow the smooth trend of ASF variations, as shown in Fig. \ref{fig:ASFdeviations}. 
Because these outliers should not be used for ASF map generation, it is essential to detect and eliminate them. 
For outlier removal, we used the filtering, based on the median absolute deviation (MAD). If the absolute deviation of a measurement exceeded the median of measurements within a 60-second time window by $2 \times \mathrm{MAD}$, the measurement was designated as an outlier and therefore excluded.

The cross-track ASF trend was subsequently obtained using a smoothing spline\cite{deBoor01}. Compared with other regression methods, such as the polynomial and local regression, the smoothing spline method has the advantage to fit the nonlinear data with a single tuning parameter.
We used the cubic smoothing spline to fit the cross-track ASF deviation data, which is a spline with piecewise third-order polynomials as follows \cite{Hastie09}:
\begin{equation} \label{eqn:cubic_splines}
  f_\mathrm{cross}(l) = \\
    \begin{cases}
      P_1(l), & l_1 \leq l < l_2 \\
      & \vdots \\
      P_{K-1}(l), & l_{K-1} \leq l < l_K
    \end{cases}	
\end{equation}
where $l$ is the cross-track distance with respect to the center of a waterway; $K$ is the total number of bias-removed measurements in the cross-track survey region; $P_k(l)$ is a piecewise third-order polynomial in $[l_k,l_{k+1}]$; $l_k$ is the cross-track distance of the $k$-th measurement location; $f_\mathrm{cross}(l)$, $f_\mathrm{cross}'(l)$, and $f_\mathrm{cross}''(l)$ are continuous in $[l_1,l_K]$; and $f_\mathrm{cross}''(l_1) = f_\mathrm{cross}''(l_K) = 0$. Furthermore, the cubic smoothing spline $f_\mathrm{cross}(l)$ was selected to minimize the following \cite{deBoor01}:
\begin{equation} \label{eqn:smoothing_splines}
  (1-p) \underbrace{\sum_{k=1}^{K} {\left ( z_k - f_\mathrm{cross}(l_k) \right )^2}}_\textrm{residual sum of squares (RSS)} + 
  p \underbrace{\int{f_\mathrm{cross}''(l)^2 dl}}_\textrm{roughness}	
\end{equation}
where $z_k$ is the cross-track ASF deviation of the $k$-th measurement and $p$ is the smoothing parameter that controls the roughness of the fitting function. The optimal smoothing parameter $p$ was chosen to minimize the leave-one-out cross-validation error \cite{Hastie09}.

\subsection{Regression Kriging with the Cross-Track ASF Trend}

\begin{figure}
  \centering
  \includegraphics[width=0.7\linewidth]{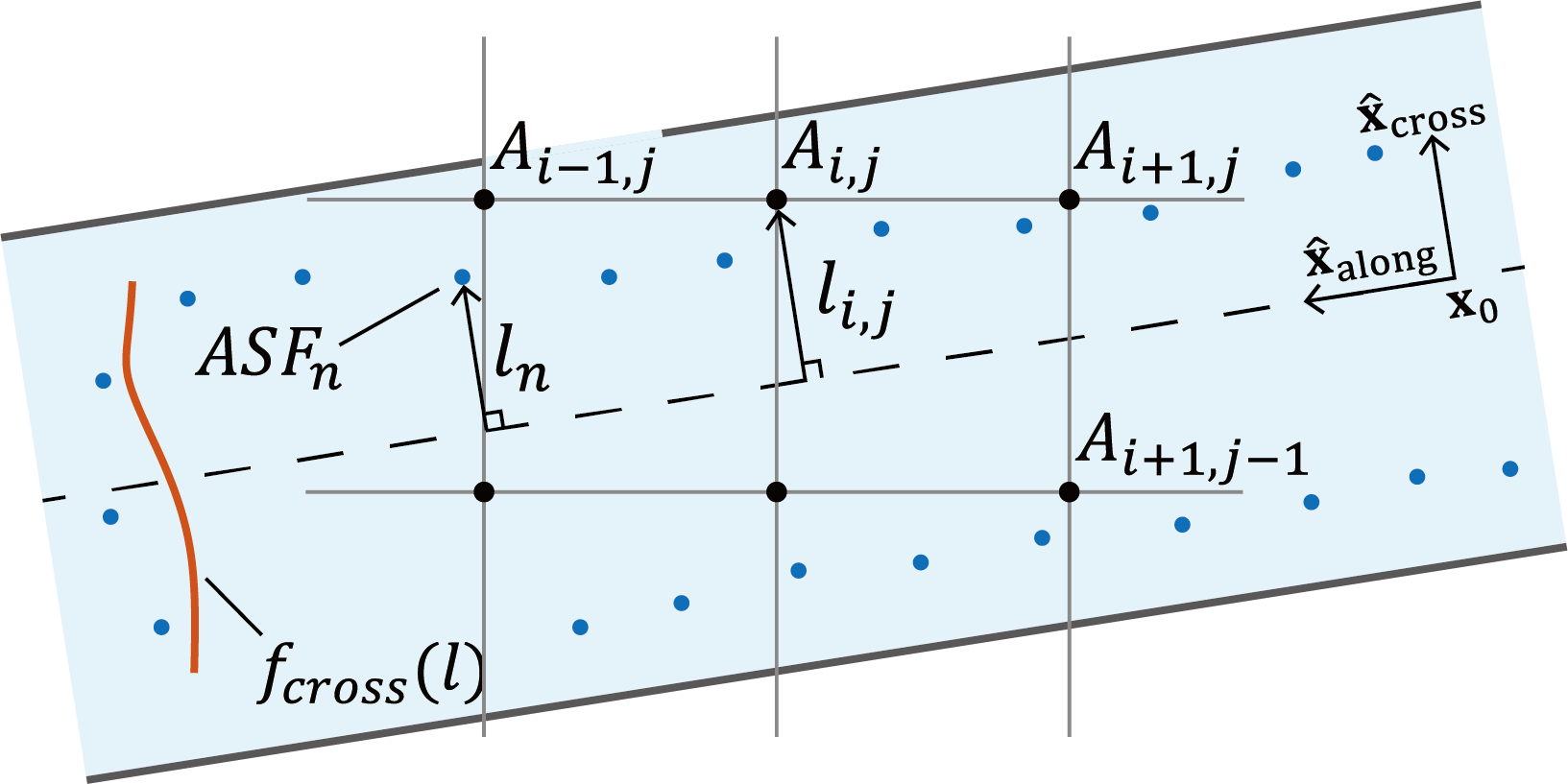}
  \caption{ASF survey locations (blue dots), grid points of ASF map (black dots), and cross-track ASF trend (red curve) in a narrow waterway.}
  \label{fig:RegKrig}
\end{figure}

The conventional method for generating ASF maps is linear interpolation \cite{RTCM17}. After the along-track and cross-track surveys as illustrated in Fig. \ref{fig:RegKrig}, the grid ASF value of an ASF map (i.e., $A_{i,j}$) can be obtained as a simple linear combination of the ASF measurements (i.e., $ASF_n$) \cite{Son19}. However, the accuracy of the grid ASF value decreases as the grid point is away from the cross-track survey area if linear interpolation is applied. For example, the accuracy of $A_{i+1,j}$ is less than that of $A_{i-1,j}$ in Fig. \ref{fig:RegKrig}.

To overcome this limitation, we propose to extract the cross-track ASF trend from a limited cross-track survey data (Section \ref{sec:ObtainAsfTrend}) and apply the trend to other areas where only a sparse along-track survey is performed. 
In a coastal area, although a universal kriging with an exponentially decaying model of coastal ASF variations has been suggested for ASF map generation \cite{Son19}, it did not provide the best results when it was applied to a narrow waterway with the obtained cross-track ASF trend.
Therefore, we propose the application of regression kriging  to generate improved ASF maps in a narrow waterway.
Unlike universal kriging, the errors caused by the regression model were not reflected in the kriging process for residuals \cite{Odeh95}. Thus, regression kriging has the advantage of minimizing the effect of regression errors (e.g., difference between the actual measurements and  fitted curve in Fig. \ref{fig:ASFdeviations}). 
The cross-track ASF data in a narrow waterway have a tendency for high regression errors even with the cubic smoothing spline model, and the order of the smoothing spline was not increased because of the overfitting problem.
We identified the problem of high regression errors during the cross-track ASF survey for the first time and proposed regression kriging as a better alternative to universal kriging to resolve this problem. 

In Fig. \ref{fig:RegKrig}, the ASF measurement $ASF_n$ at the $n$-th measurement location can be expressed using the cross-track ASF trend, $f_\mathrm{cross}$, as:
\begin{equation}
\begin{split}
    \label{eqn:ASFn}
    ASF_n = \mu_0 + f_\mathrm{cross}(l_n) + \varepsilon(\mathbf{x}_n) \\
    \mu_0 = \frac{1}{N} \sum_{n=1}^N \left ( ASF_n-f_\mathrm{cross}(l_n) \right)
\end{split}
\end{equation}
where $\mu_0$ is the detrended average of the ASF measurements, $\mathbf{x}_n$ is the position vector of the $n$-th measurement location, $l_n$ is the cross-track distance of $\mathbf{x}_n$ from the center of the waterway, $N$ is the number of ASF measurements, and $\varepsilon(\mathbf{x}_n)$ is the remaining stochastic term that is defined as the residual of the $n$-th ASF measurement.
Subsequently, the grid ASF value, $A_{i,j}$, obtained by regression kriging is expressed as follows:
\begin{equation}
    \label{eqn:regression_krig}
    A_{i,j} = \mu_0 + f_\mathrm{cross}(l_{i,j})+\sum^N_{n=1}{w_n\varepsilon(\mathbf{x}_n)}
\end{equation}
where $l_{i,j}$ is the cross-track distance of the grid point $(i,j)$ from the center of the waterway, and $w_n$ is the weight for the $n$-th residual obtained by \eqref{eqn:ordinary_kriging}.
\begin{equation}
    \label{eqn:ordinary_kriging}
    \resizebox{.9\linewidth}{!}{$
        \begin{bmatrix}
            w_1\\\vdots\\w_N\\\lambda
        \end{bmatrix}
        =
        {
        \begin{bmatrix}
            \gamma(\mathbf{x}_1,\mathbf{x}_1) & \dots &  \gamma(\mathbf{x}_1,\mathbf{x}_N) & 1\\
            \vdots & \ddots & \vdots & \vdots\\
            \gamma(\mathbf{x}_N,\mathbf{x}_1) & \dots &  \gamma(\mathbf{x}_N,\mathbf{x}_N) & 1\\
            1 & \dots & 1 & 0
        \end{bmatrix}
        }^{-1}
        \begin{bmatrix}
            \gamma(\mathbf{x}_1,\mathbf{x}_{i,j})\\\vdots\\\gamma(\mathbf{x}_N,\mathbf{x}_{i,j})\\1
        \end{bmatrix}
    $}
\end{equation}
where $\lambda$ is the Lagrange multiplier, $\mathbf{x}_{i,j}$ is the position vector of the grid point $(i,j)$, and $\gamma(\mathbf{x}_a,\mathbf{x}_b)$ is the semivariance between the two positions $\mathbf{x}_a$ and $\mathbf{x}_b$. 

It is useful to compare the grid ASF value, $A_{i,j}$, obtained from regression kriging in (\ref{eqn:regression_krig}) and universal kriging in (\ref{eqn:universal_krig}).
\begin{equation}
    \label{eqn:universal_krig}
    A_{i,j} = \sum^N_{n=1}{w_n ASF_n}
\end{equation}
\begin{equation*}
    \resizebox{1.0\linewidth}{!}{$
        \begin{bmatrix}
            w_1\\\vdots\\w_N\\\lambda_0\\\lambda_1
        \end{bmatrix}
        = 
        {
        \begin{bmatrix}
            \gamma(\mathbf{x}_1,\mathbf{x}_1) & \dots &  \gamma(\mathbf{x}_1,\mathbf{x}_N) & 1 & f_\mathrm{cross}(l_1)\\
            \vdots & \ddots & \vdots & \vdots & \vdots\\
            \gamma(\mathbf{x}_N,\mathbf{x}_1) & \dots &  \gamma(\mathbf{x}_N,\mathbf{x}_N) & 1 & f_\mathrm{cross}(l_N)\\
            1 & \dots & 1 & 0 & 0\\
            f_\mathrm{cross}(l_1) & \dots & f_\mathrm{cross}(l_N) & 0 & 0\\
        \end{bmatrix}
        }^{-1}
        \begin{bmatrix}
            \gamma(\mathbf{x}_1,\mathbf{x}_{i,j})\\\vdots\\\gamma(\mathbf{x}_N,\mathbf{x}_{i,j})\\1\\f_\mathrm{cross}(l_{i,j})
        \end{bmatrix}
    $}
\end{equation*}
In universal kriging, the cross-track ASF trend, $f_\mathrm{cross}$, is used as an additional constraint when calculating the weights, and $A_{i,j}$ is represented by the weighted combination of $ASF_n$. 
It should be noted that the expressions of (\ref{eqn:ASFn})--(\ref{eqn:universal_krig}), which include $f_\mathrm{cross}$, are our novel implementations to reflect the cross-track ASF trend for the ASF map generation, although the generic paradigm of universal kriging and regression kriging already exist. 

\section{Field Test Results} 

\begin{figure}
  \centering
  \includegraphics[width=0.9\linewidth]{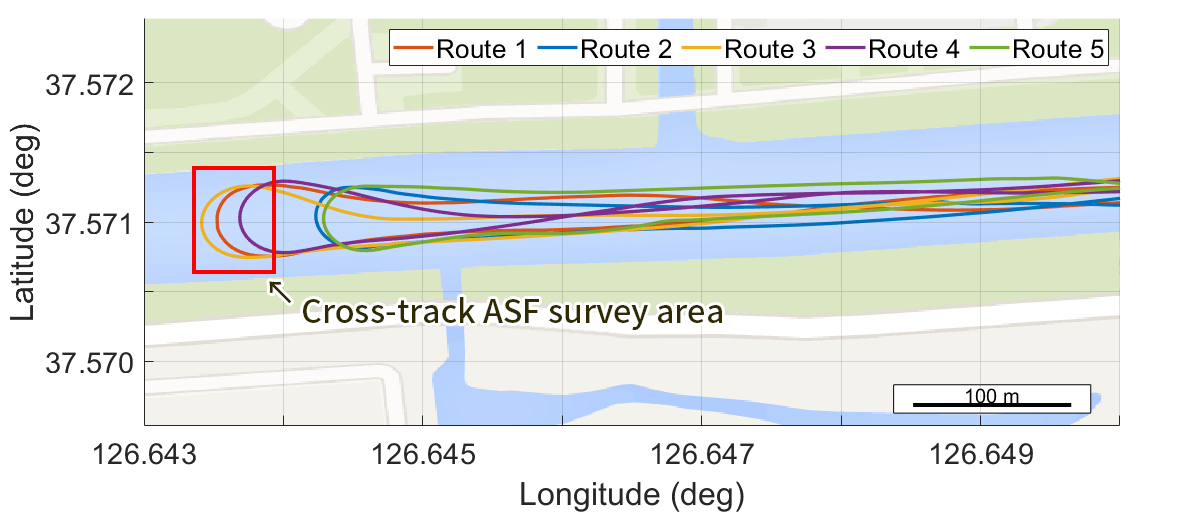}
  \caption{Part of the field test area in the Ara Waterway, Incheon, Korea. Data collected along Route 1 were used for the ASF map generation, and the positioning performance was evaluated along the remaining routes.}
  \label{fig:SurveyMap}
\end{figure}

To evaluate the positioning accuracy enhancement from the proposed ASF map generation methods, we conducted an ASF survey and performance tests on the Ara Waterway located in Incheon, Korea. The eLoran ASF values, time of arrival (TOA) measurements, and ground truth positions were collected along five routes. A small portion of the routes is shown in Fig. \ref{fig:SurveyMap}. The ASF maps were generated using the data along Route 1 (2.4 km) based on the three different methods: linear interpolation, universal kriging, and regression kriging. The grid size of the ASF maps was set to 100 m.

\begin{figure}
  \centering
  \includegraphics[width=0.9\linewidth]{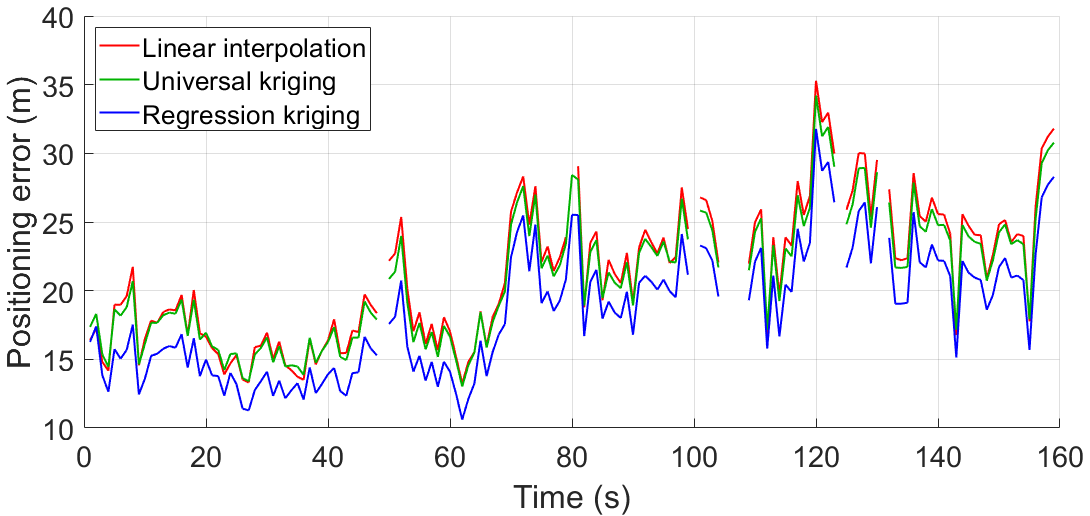}
  \caption{Comparison of the 2-D RMS positioning errors according to the three different ASF map generation methods. A small portion of the test data along Route 3 in Fig. \ref{fig:SurveyMap} is presented.}
  \label{fig:PosError}
\end{figure}

The positioning accuracy was evaluated along the remaining routes, that is, Routes 2 through 5 (10.7 km in total), which were not used for the ASF map generation. The position solutions were calculated using the TOA measurements and  generated ASF maps. Fig. \ref{fig:PosError} compares the  positioning errors according to each ASF map generation method along a small portion of Route 3 as an example. When the ASF maps generated by the proposed regression kriging were utilized, the positioning errors were smaller than those in the other cases. 
The 2-D root-mean-square (RMS) positioning errors, when linear interpolation, universal kriging, and regression kriging were utilized for the ASF map generation, were 19.56, 19.02, and 17.94 m, respectively, along the 10.7-km test routes. 

The cross-track ASF survey does not necessarily guarantee a high performance enhancement. 
The conventional method to generate ASF maps relies on linear interpolation and does not perform the cross-track surveys. If this conventional approach without a cross-track survey was applied, the 2-D RMS positioning error was 19.95 m along the same test routes, which was only 0.39 m less than the case of linear interpolation with a cross-track survey.
However, our approach demonstrated an error of approximately 2 m less than that of the conventional approach.
Therefore, the advantage of the cross-track ASF survey was not evident until the proposed regression kriging with the cross-track ASF trend was applied.

\section{Conclusion}

South Korea has been actively developing eLoran as a complementary navigation system to GNSSs and recently announced the operation of its eLoran testbed system.
Demonstrating the satisfactory performance of eLoran in a testbed is an important step towards the nationwide rollout of the Korean eLoran system.
In this letter, we demonstrate the Korean eLoran accuracy in a narrow waterway for the first time. To overcome the practical difficulty of surveying a narrow waterway, a simple cross-track survey was proposed. To effectively utilize the cross-track survey data for ASF map generation, we proposed methods to extract the cross-track ASF trends and apply regression kriging. 
The proposed methods demonstrated an improvement in the eLoran positioning accuracy in the field tests along the 10.7-km test routes as compared to the existing methods.

\ifCLASSOPTIONcaptionsoff
  \newpage
\fi




\bibliographystyle{./IEEEtran}
\bibliography{./IEEEabrv,./mybibfile}

\end{document}